\documentclass[reprint, superscriptaddress, amsmath,amssymb, aps,prd, preprintnumbers]{revtex4-2}   
\RequirePackage[colorlinks=true
,urlcolor=magenta
,anchorcolor=blue
,citecolor=magenta
,filecolor=blue
,linkcolor=blue
,menucolor=blue
,linktocpage=true
,pdfproducer=medialab
,pdfa=true
]{hyperref}

\usepackage{amsmath,amssymb,amsthm,amsfonts}
\usepackage{cprotect}
\usepackage{graphicx,tabularx}
\usepackage{float}
\usepackage{multirow}  
\usepackage{cancel}
\usepackage{comment}
\usepackage{physics}
\usepackage{array}
\usepackage{enumitem}
\newcolumntype{P}[1]{>{\centering\arraybackslash}p{#1}}
\usepackage{natbib}
\usepackage{booktabs}

\usepackage{xcolor}
\usepackage{dcolumn}
\usepackage{bm}

\usepackage{bm,epstopdf,natbib,hyperref,color,verbatim,multirow,bm,tikz,xcolor}
\hypersetup{colorlinks=true,urlcolor=blue,citecolor=blue,linkcolor=blue,menucolor=blue,anchorcolor=blue,filecolor=blue}
\everymath{\displaystyle}
\definecolor{lime}{HTML}{A6CE39}
\DeclareRobustCommand{\orcidicon}{
	\begin{tikzpicture}
	\draw[lime, fill=lime] (0,0) 
	circle [radius=0.16] 
	node[white] {{\fontfamily{qag}\selectfont \tiny ID}};
	\draw[white, fill=white] (-0.0625,0.095) 
	circle [radius=0.007];
	\end{tikzpicture}
	\hspace{-3mm}
}
\foreach \x in {A, ..., Z}{\expandafter\xdef\csname orcid\x\endcsname{\noexpand\href{https://orcid.org/\csname orcidauthor\x\endcsname}
			{\noexpand\orcidicon}}
}

\begin{document}
\preprint{LHAASO-MSA-25-0103}
\title{Energy-Dependent Shifts of Medium-Scale Anisotropies in Very-High-Energy Cosmic Rays Observed by LHAASO-KM2A}
 
\author{Zhen Cao}
\affiliation{Key Laboratory of Particle Astrophysics \& Experimental Physics Division \& Computing Center, Institute of High Energy Physics, Chinese Academy of Sciences, 100049 Beijing, China}
\affiliation{University of Chinese Academy of Sciences, 100049 Beijing, China}
\affiliation{TIANFU Cosmic Ray Research Center, Chengdu, Sichuan,  China}
 
\author{F. Aharonian}
\affiliation{TIANFU Cosmic Ray Research Center, Chengdu, Sichuan,  China}
\affiliation{University of Science and Technology of China, 230026 Hefei, Anhui, China}
\affiliation{Yerevan State University, 1 Alek Manukyan Street, Yerevan 0025, Armenia}
\affiliation{Max-Planck-Institut for Nuclear Physics, P.O. Box 103980, 69029  Heidelberg, Germany}
 
\author{Y.X. Bai}
\affiliation{Key Laboratory of Particle Astrophysics \& Experimental Physics Division \& Computing Center, Institute of High Energy Physics, Chinese Academy of Sciences, 100049 Beijing, China}
\affiliation{TIANFU Cosmic Ray Research Center, Chengdu, Sichuan,  China}
 
\author{Y.W. Bao}
\affiliation{Tsung-Dao Lee Institute \& School of Physics and Astronomy, Shanghai Jiao Tong University, 200240 Shanghai, China}
 
\author{D. Bastieri}
\affiliation{Center for Astrophysics, Guangzhou University, 510006 Guangzhou, Guangdong, China}
 
\author{X.J. Bi}
\affiliation{Key Laboratory of Particle Astrophysics \& Experimental Physics Division \& Computing Center, Institute of High Energy Physics, Chinese Academy of Sciences, 100049 Beijing, China}
\affiliation{University of Chinese Academy of Sciences, 100049 Beijing, China}
\affiliation{TIANFU Cosmic Ray Research Center, Chengdu, Sichuan,  China}
 
\author{Y.J. Bi}
\affiliation{Key Laboratory of Particle Astrophysics \& Experimental Physics Division \& Computing Center, Institute of High Energy Physics, Chinese Academy of Sciences, 100049 Beijing, China}
\affiliation{TIANFU Cosmic Ray Research Center, Chengdu, Sichuan,  China}
 
\author{W. Bian}
\affiliation{Tsung-Dao Lee Institute \& School of Physics and Astronomy, Shanghai Jiao Tong University, 200240 Shanghai, China}
 
\author{A.V. Bukevich}
\affiliation{Institute for Nuclear Research of Russian Academy of Sciences, 117312 Moscow, Russia}
 
\author{C.M. Cai}
\affiliation{School of Physical Science and Technology \&  School of Information Science and Technology, Southwest Jiaotong University, 610031 Chengdu, Sichuan, China}
 
% ...
\collaboration{The LHAASO Collaboration}

\email{hejy@pmo.ac.cn}
\email{zhaosp@pmo.ac.cn}
\email{zhangyi@pmo.ac.cn}
\email{yuanq@pmo.ac.cn}
\email{bianwenyi@sjtu.edu.cn}
\email{lidan@ihep.ac.cn}
\email{gwenael.giacinti@sjtu.edu.cn}


\begin{abstract}

Small deviations from isotropy in the arrival directions of Galactic cosmic rays serve as a unique probe of the local magnetic environment. In this Letter, we report observations of medium-scale anisotropies at energies above 10 TeV using the LHAASO-KM2A array. Our analysis identifies four regions of excess and four regions of deficit, each spanning angular scales of approximately ten degrees. Crucially, we detect significant energy-dependent shifts in the centroids of two excess regions: Region B and the newly identified Region $\mathrm{\widetilde{D}}$. We also characterize the energy evolution of the fractional relative intensity across both excess and deficit regions. These findings imply that the observed anisotropies are shaped by the specific realization of the local turbulent magnetic field within the cosmic ray scattering length. Such energy-dependent behaviors impose strict constraints on local turbulence models and cosmic ray propagation theories.

\end{abstract}

\maketitle


\noindent{\bf Introduction.}

In the presence of a turbulent magnetic field, galactic cosmic rays (CR) are expected to lose their initial directional information after prolonged propagation, resulting in a nearly isotropic flux. However, numerous ground-based experiments have identified anisotropies with amplitudes ranging from \(10^{-4}\) to \(10^{-3}\) in the very-high-energy range \citep{ASgamma_LSA2006, Milagro_LSA2009, ARGO_LSA2018, HAWC_LSA2018, IceCube_MediumScale2011, IceCube_LSA2012, IceTop_LSA2013, IceCube_2016}. On the large scale, this anisotropy is generally characterized by a dipole whose amplitude and phase are energy-dependent. The origins of this large-scale anisotropy remain poorly understood. Current hypotheses suggest that it may be due to a combination of several factors, including the distribution of CR sources, the diffusion process in the galaxy, nearby sources, and the local magnetic field near the solar system \citep{Blasi_2012,Kumar_2014,Battaner_2015,PhysRevLett.114.021101,PhysRevLett.117.151103,Qiao_2023}.

Beyond the large-scale structure, medium-scale excess regions have been observed down to an angular size of \(\sim 10^{\circ}\), as reported by ground-based experiments in the northern hemisphere, such as Tibet AS$\gamma$ \citep{ASgamma_MediumScale2007}, Milagro \citep{Milagro_MediumScale2009}, ARGO-YBJ \citep{ARGO_MediumScale2013}, HAWC \citep{HAWC_2014, HAWC_LSA2018}, and GRAPES-3 \citep{GRAPES3_MediumScale2024}, and in the southern hemisphere by IceCube \citep{IceCube_MediumScale2011}. Notable excess structures in the northern sky are referred to as Region A ($\alpha \sim 60^{\circ}$, $\delta \sim -10^{\circ} \sim 10^{\circ}$) and Region B ($\alpha \sim 120^{\circ}$, $\delta \sim 40^{\circ}$), which correspond to Regions 1 and 2 in ARGO-YBJ data, respectively \citep{ARGO_MediumScale2013}. Additionally, a hot spot known as Region C ($\alpha \sim 200^{\circ}$, $\delta \sim 20^{\circ}$) as seen in HAWC data \citep{HAWC_LSA2018} corresponds to Region 4 ($\alpha \sim 210^{\circ}$, $\delta \sim 30^{\circ}$) in ARGO-YBJ results \citep{ARGO_MediumScale2013}. Both ARGO-YBJ and HAWC have discovered new and distinct structures, such as Region D ($\alpha \sim 5^{\circ}$, $\delta \sim 40^{\circ}$) in HAWC \citep{HAWC_LSA2018} and an elongated structure, Region 3 ($\alpha \sim 240^{\circ}$, \(15^{\circ} < \delta < 55^{\circ}\)), in ARGO-YBJ \citep{ARGO_MediumScale2013}.

While the dipole anisotropy is predicted in the diffusion scenario, the origin of the medium-scale anisotropy is debated. A nearby source like the Geminga pulsar is thought to be a solution for the excess CRs \citep{NearbySource2008}, and CR acceleration from magnetic reconnection in the magnetotail is also considered as a possible source of the energetic particle excess \citep{MagneticReconnection2010}. Alternative explanations include particles escaping from a confined source region through a magnetically confined loss cone \citep{MagneticMirror2008}, a nearby field configuration acting as a magnetic lens \citep{MagneticLens2015,Schwadron_2014}, and CRs deflected by an anisotropic Alfvén wave spectrum \citep{PreferrDirection2010}. 
Besides, the re-distribution of anisotropic CRs due to scattering on perturbations induced in the local interstellar magnetic field by the heliosphere wake \citep{NonisotropicTransport_2013}, or the electric field structure of the outer heliosphere \citep{ElectricField2013} can be used to interpret the observed medium-scale structures. 
Some exotic models have been proposed, such as the emission of strangelets during the transition of a neutron star to a more compact star \citep{Strangelet2013}, or as a result of internal transitions involving nuclear deconfinement of the quark content within a neutron star \citep{NeutronStar2014}. These strangelets, once neutralized in nearby molecular clouds, can propagate rectilinearly towards Earth. Furthermore, it is proposed that the medium-scale anisotropies are the natural results of propagation of CRs through the turbulent magnetic fields around the Earth in\citep{Giacinti2012,Ahlers2014,Ahlers_2015,Kuhlen_2022,Zhang_2024}.


The measurements of previous experiments primarily focus on the fractional relative intensity and angular power spectra, and the morphology of the excess regions has not been measured. However, the morphology of the medium-scale structures may encode the information from local turbulence magnetic fields. As shown in the studies of \cite{Giacinti2012,Schwadron_2014}, the resulting anisotropy patterns are energy-dependent and arise naturally from the specific local realization of the turbulent magnetic field within the cosmic ray scattering length. In this Letter, with the data of CRs accumulated by the Large High Altitude Air Shower Observatory (LHAASO), we measure the medium-scale anisotropies above 10 TeV and investigate the energy-dependent locations' evolution. Additional details on energy reconstruction, comparison with prior experiments, anti-/extended-sidereal time analysis, and numerical simulations are provided in the Supplemental Material \cite{SupplementalMaterial}. 


\noindent{\bf Experiment.}
LHAASO (100.01°E, 29.35°N) is a ground-based air shower observatory located at an altitude of 4,410 meters in Daocheng, China \cite{Aharonian_2021}. It is a large, hybrid extensive air shower array comprising three sub-arrays: a 1.3 km$^2$ array (KM2A), a 78,000 m$^2$ water Cherenkov detector array (WCDA), and 18 wide field-of-view air Cherenkov/fluorescence telescopes (WFCTA). The KM2A consists of 5,216 electromagnetic detectors (EDs) and 1,188 muon detectors (MDs). LHAASO-KM2A has been operating in full array mode since July 20, 2021. This study utilizes data from KM2A collected over a three-year period, from July 20, 2021, to July 19, 2024.

\noindent{\bf Data Analysis.}
We utilized the official simulation chain to estimate the angular and energy resolution of our study. CORSIKA (version 77410) \cite{Corsikaguide} was employed as the air shower simulator. For high-energy interactions, we used the models of QGSJETII-04 \cite{QGSJET}, while for low-energy interactions, the FLUKA code \cite{Fluka} was utilized. The CR particles were categorized into five groups: protons, helium, CNO, MgAlSi, and iron. The Geant4 toolkit \cite{Geant4} with the G4KM2A module \cite{G4KM2A} was used to simulate the detector response.

Both Monte Carlo (MC) and experimental samples were selected based on the following criteria: (1) the reconstructed zenith angle must be less than \(40^{\circ}\); (2) at least 20 electromagnetic detectors (EDs) are used for reconstruction; (3) the reconstructed shower core must be located within the array; (4) events must have a reconstructed energy above 10\,TeV.

To estimate the primary energy of a CR event, a combination parameter (\(N_{e,\mu} \sim N_{e} + 2.8N_{\mu}\)) was employed, where \(N_{e}\) is the number of recorded electromagnetic particles and \(N_{\mu}\) is the number of recorded muons \cite{LHAASO_APSED}.  Events are divided into five energy bins, each with a width of \(\Delta \log_{10} E = 0.25\), with energy expressed in TeV. Table \ref{tab:binning} summarizes the median energy (which is determined by MC data weighted with the H3a model \cite{gaisser2013cosmic}) and corresponding statistics for each energy bin. The energy dispersion (defined as the half-width of the 68\% containment interval) is 16\% at 16.5 TeV (the first bin) and improves to 10\% at 69.9 TeV (the fourth bin). The corresponding LHAASO-KM2A angular resolutions at these energies are $0.8^{\circ}$ and $0.3^{\circ}$, respectively.

\begin{table}[!htbp]
\centering
\begin{tabular}{ccc}
\toprule
$\log_{10}(E_\text{rec}/\mathrm{TeV})$ & Median (TeV) & Events ($\times 10^{10}$)\\ 
\midrule
1.00 -- 1.25  & 16.5 & 1.65 \\
1.25 -- 1.50  & 23.8 & 4.19 \\
1.50 -- 1.75  & 39.8 & 3.37 \\
1.75 -- 2.00  & 69.9 & 1.54 \\
$>$ 2.00  & 159.2 & 1.00 \\
\bottomrule
\end{tabular}
\caption{Statistics and median energy for each analysis bin.}
\label{tab:binning}
\end{table}

To estimate reference maps of CRs, the all-distance equi--zenith angle method was used to calculate the relative intensity in each arrival direction bin \cite{amenomori2005northern}. This method has been widely applied in previous ground-based experiments \citep{ASgamma_LSA2006,ASgamma_LSA2017,ARGO_LSA2018}. This method utilizes events within the equi-zenith angle belt, excluding those in the on-source bin, as off-source events \cite{amenomori2005northern}. The $\chi^2$ is constructed as follows:
\begin{equation}
    \chi ^{2}= \sum_{i,j,k} \frac{\left( \frac{N(\theta_{i},\phi_{j},\tau_{k})}{I(\alpha_{\rho},\delta_{\nu})} - \frac{1}{n_{\theta_{i}}-1}\sum_{j'\neq j}\frac{N(\theta_{i},\phi_{j'},\tau_{k})}{I(\alpha_{\rho'},\delta_{\nu'})}\right)^{2}}{\sigma^{2} _{i,j,k}},
\label{eq:chi2_eq}
\end{equation}
where the relative intensity \(I(\alpha,\delta)\) is defined as the ratio of observed events \(N\) to isotropic background \(B\), i.e., \(I(\alpha,\delta) = N(\alpha,\delta)/B(\alpha,\delta)\). Events are binned by their local sidereal time (\(\tau_{k}\)), and further divided into rings based on zenith angle (\(\theta_{i}\)) in local horizontal coordinates. The azimuthal pixels (\(\phi_{j}\)) are designed to ensure identical solid angles, with the number of pixels in the \(i\)-th zenith ring denoted as \(n_{\theta_{i}}\). The event count in the direction \((\theta_{i},\phi_{j})\) for a given time interval \(\tau_{k}\) is denoted as \(N(\theta_{i},\phi_{j},\tau_{k})\). Each pixel, defined in local horizontal coordinates \((\theta_{i},\phi_{j},\tau_{k})\), can be converted to equatorial coordinates \((\alpha_{\rho},\delta_{\nu})\). The term \(\sigma^{2}_{i,j,k} \approx \frac{N(\theta_{i},\phi_{j},\tau_{k})}{I^{2}(\alpha_{\rho},\delta_{\nu})}\) represents the uncertainty of the numerator. By iteratively minimizing the $\chi^2$ as outlined in \citet{amenomori2005northern}, sky maps of relative intensity \(I(\alpha,\delta)\) are derived. \(I(\alpha,\delta)\) is normalized for all declinations, which results in the loss of the $m=0$ component in our measurements \citep{Ahlers_2016}. Nevertheless, this method remains capable of efficiently reconstructing the $m\neq0$ components, ensuring the validity of the spherical harmonic expansion for \(I(\alpha,\delta)\) in the following analysis.

\begin{figure}[!htbp]
\centering
\includegraphics[width=0.4\textwidth]{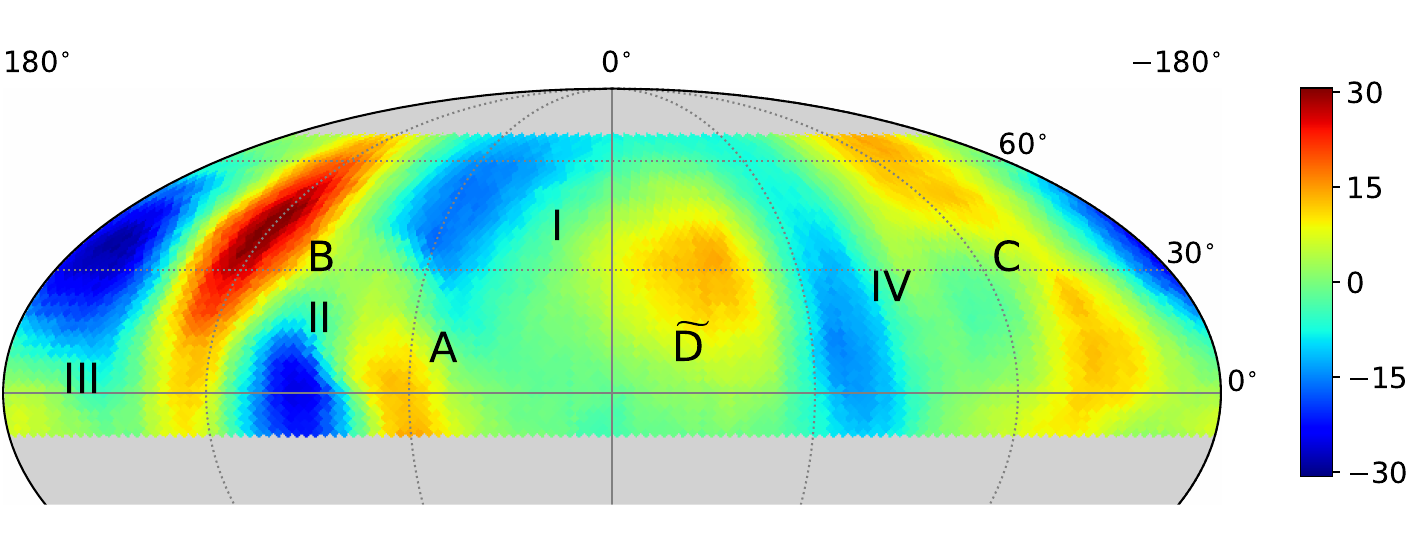}
\caption{Significance sky map of medium-scale anisotropy in equatorial coordinates with a median energy of $\sim35\,\mathrm{TeV}$. The map is smoothed with an $18^\circ$ radius top-hat function. The excess regions A, B, C, and $\mathrm{\widetilde{D}}$ exhibit significances of $14.2\,\sigma$, $30.6\,\sigma$, $14.4\,\sigma$, and $14.3\,\sigma$, respectively, while the deficit regions I, II, III, and IV show significances of $-16.7\,\sigma$, $-25.1\,\sigma$, $-29.1\,\sigma$, and $-15.1\,\sigma$.}
\label{fig:ts}
\end{figure}

The sky maps of \(I(\alpha,\delta)\) contain all the information regarding the angular power spectrum; however, medium-scale anisotropies are associated with the angular power spectrum for \(l \geq 3\). Given that the field of view of LHAASO is not full-sky, to extract smaller-scale information, it is necessary to remove larger-scale structures corresponding to the lower multipoles. The intensity \(\delta I(\alpha,\delta) = I(\alpha,\delta) - 1\) is expanded using a real spherical harmonic basis \cite{IceCube_MediumScale2011,HAWC_LSA2018}, such that \(\delta I(\alpha,\delta) = \delta I(\alpha,\delta)_{\ell \leq 2} + \delta I(\alpha,\delta)_{\geq 3}\). The component \(\delta I(\alpha,\delta)_{\ell \leq 2}\) is represented as follows:

\begin{equation}
\begin{split}
    \delta I(\alpha,\delta)_{\ell \leq2} &=  m_{0} + p_{y} \cos\delta \sin\alpha + p_{z} \sin\delta +  p_{x} \cos\delta \cos\alpha \\
    &+ Q_{1} \cos^{2}\delta \sin 2\alpha  + Q_{2} \sin 2\delta \sin \alpha \\ 
    &+ \frac{1}{2}Q_{3}( 3 \sin^{2} \delta - 1 ) \\
    &+ Q_{4} \sin 2\delta \cos \alpha + Q_{5} \cos^{2}\delta \cos 2\alpha,
\end{split}
\label{eq:expand}
\end{equation}
where \(m_0\) is the monopole component, \((p_x, p_y, p_z)\) are the dipole moment components, and \((Q_1, Q_2, Q_3, Q_4, Q_5)\) are the components of the quadrupole moment. The medium-scale and smaller-scale anisotropies are characterized by:
\begin{equation}
I(\alpha,\delta)_{\ell \geq 3} = I(\alpha,\delta) - \delta I(\alpha,\delta)_{\ell \leq2},
\label{eq:imsa}
\end{equation}
where \(\delta I(\alpha,\delta)_{\ell \leq 2}\) is determined by fitting Equation \ref{eq:expand} to the map of \(I(\alpha,\delta)\). The free parameters in this fit are the coefficients of the multipoles, specifically \((m_0, p_x, p_y, p_z, Q_1, Q_2, Q_3, Q_4, Q_5)\). The quantity \(I(\alpha,\delta)_{\ell \geq 3}\) represents the measurement of interest.


\begin{figure}[!htbp]
\centering
\includegraphics[width=0.4\textwidth]{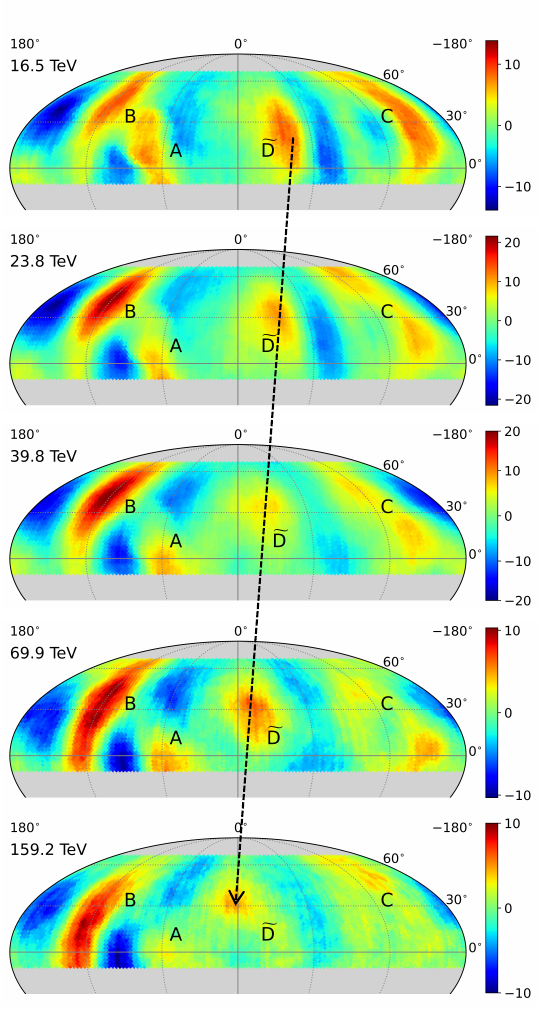}
\caption{Significance sky maps of medium-scale anisotropy in equatorial coordinates. Each map is smoothed with an $18^\circ$  top-hat function. The maps in each row correspond to the bins in Table \ref{tab:binning} with median energies of 16.5 TeV, 23.8 TeV, 39.9 TeV, 69.9 TeV, and 159.2 TeV. The dashed line shows the evolution trend for Region $\mathrm{\widetilde{D}}$.}
\label{fig:skymap}
\end{figure}

\noindent{\bf Results.}

The LHAASO-KM2A significance maps of medium-scale anisotropies, for all combined bins with a median energy of approximately 35 TeV, are depicted in Figure \ref{fig:ts}, where the significance is
calculated using the method of Li and Ma \cite{Li-Ma_1983}. The plot highlights four prominent excess regions, labeled as Region A, B, C, and $\mathrm{\widetilde{D}}$ for convenience (the label $\mathrm{\widetilde{D}}$ is used to distinguish this from the Region D observed by HAWC). 

Both Region A (\(\alpha \approx 60^{\circ}\)) and Region B (\(\alpha \approx 120^{\circ}\)) have been previously identified by other ground-based experiments located in the northern hemisphere \cite{Milagro_MediumScale2009,ARGO_MediumScale2013,HAWC_2014,HAWC_LSA2018,GRAPES3_MediumScale2024}. The significance levels for Regions A and B are 14.2\(\,\sigma\) and 30.6\(\,\sigma\), respectively. Region C exhibits an irregular shape with two sub-regions: one at \(\alpha \approx 210^{\circ}\), \(\delta \approx 15^{\circ}\), coinciding with Region 4 of ARGO-YBJ (or Region C of HAWC), and another at \(\alpha \approx 240^{\circ}\), \(\delta \approx 60^{\circ}\), matching Region 3 of ARGO-YBJ. The maximum significance of Region C in our map reaches approximately 14.4 \(\sigma\). 
Region $\mathrm{\widetilde{D}}$ (\(\alpha \approx 330^{\circ}\)) in our sky map, with a significance of 14.3 \(\sigma\), has not been previously reported by other experiments. Its position differs from HAWC's Region D, which is most significant at 2 TeV (located around \(\alpha \approx 5^{\circ}\)) and diminishes at higher energies. Figure \ref{fig:ts} also identifies four significant deficit regions: Region I (\(-16.7\, \sigma\)), Region II (\(-25.1\, \sigma\)), Region III (\(-29.1\, \sigma\)), and Region IV (\(-15.1\, \sigma\)).

The results for medium-scale anisotropy across different energy intervals are presented in Figure \ref{fig:skymap}. The location of excess regions, particularly Region $\mathrm{\widetilde{D}}$, appears to be energy-dependent, with its right ascension shifting eastward as energy increases. Region C, however, exhibits a complex morphological structure that splits into two sub-regions above 24 TeV. Since this morphology cannot be adequately described by a simple two-dimensional Gaussian distribution, the energy evolution of its centroid is not discussed in this work. To quantify the energy-dependent shift in other features, all other regions labeled in Figure \ref{fig:skymap} were modeled using a two-dimensional Gaussian function to obtain the fitted central locations. Our findings indicate a significant correlation between energy and Region B's declination, as well as Region $\widetilde{\text{D}}$'s right ascension.

\begin{figure}[!htbp]
\centering
\includegraphics[width=0.45\textwidth]{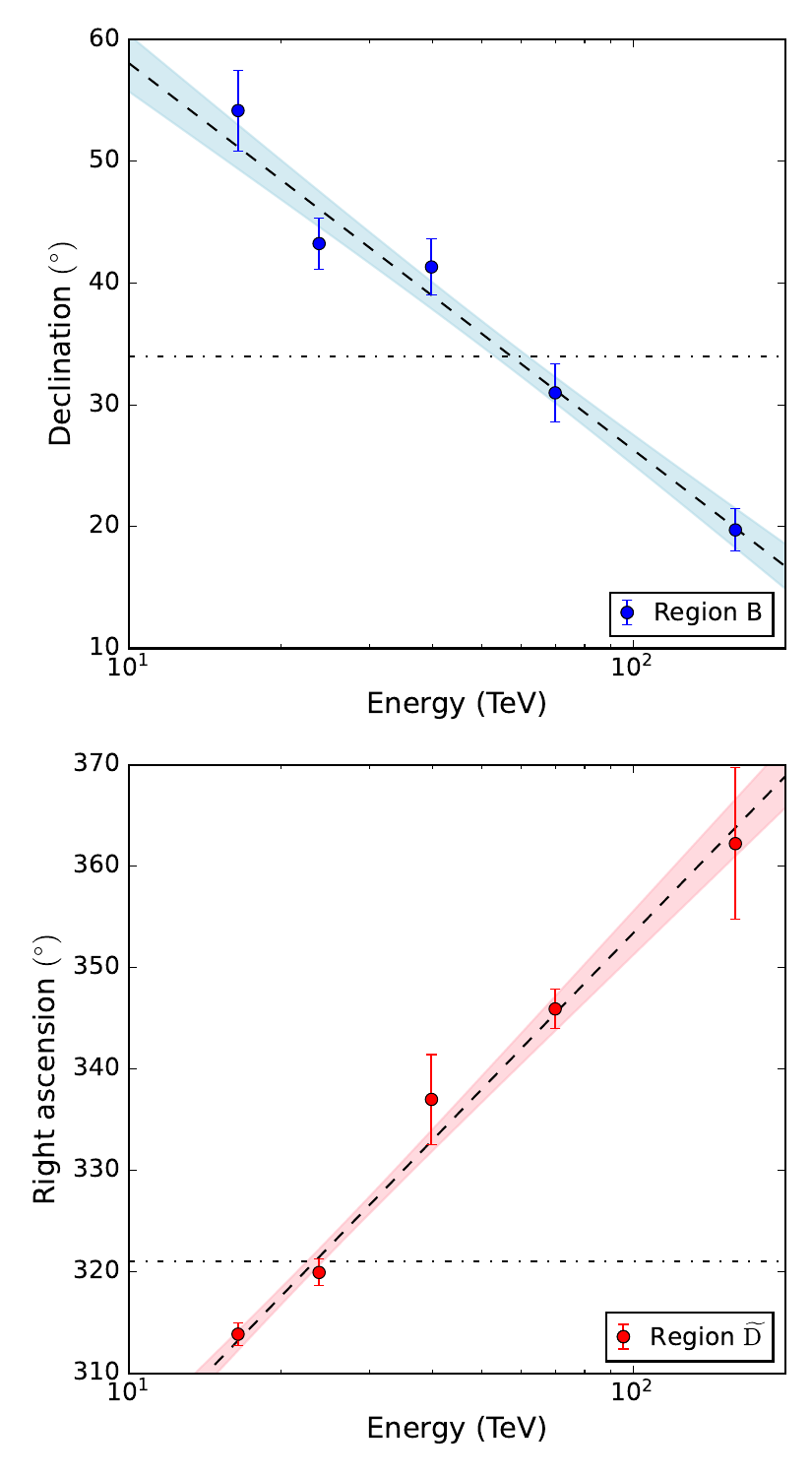}
\caption{Energy-dependent positions of Region B and Region $\mathrm{\widetilde{D}}$. The vertical axis represents declination for Region B and right ascension for Region $\mathrm{\widetilde{D}}$. The error bars indicate the total uncertainty, derived from the quadratic sum of the statistical and systematic components. The dashed-dotted lines are the best fit of the 0th-order polynomial fitting. The dashed lines and light blue/red areas represent the best fit and the 68\% confidence level for 1st-order polynomial fitting, respectively. For Region B, \(\chi^{2}_{0} = 135.6\) and \(\chi^{2}_{1} = 3.6\); for Region $\mathrm{\widetilde{D}}$, \(\chi^{2}_{0} = 245.8\) and \(\chi^{2}_{1} = 2.6\).}
\label{fig:Eposition}
\end{figure}

Figure \ref{fig:Eposition} illustrates the fitted central positions of Region B and Region $\mathrm{\widetilde{D}}$ as a function of energy. The error bars for each energy bin represent the quadratic sum of statistical and systematic uncertainties. These systematic uncertainties primarily originate from the choice of background estimation method and the fitting model. Regarding background estimation, we used both the direct integration method incorporating a maximum-likelihood algorithm \citep{Ahlers_2016} and the all-distance equi-zenith angle method \citep{amenomori2005northern} adopted in this work. For the morphological fitting, while our primary analysis utilizes a two-dimensional Gaussian distribution to determine the centroid, we also tested alternative models, such as independent one-dimensional Gaussian fits for right ascension and declination. The systematic error is defined as the deviation in results obtained from these different background estimation techniques and fitting models. The upper plot shows Region B's declination decreasing with energy, while the lower plot illustrates Region $\mathrm{\widetilde{D}}$'s right ascension increasing. No significant movement is observed in Region B's right ascension or Region $\mathrm{\widetilde{D}}$'s declination. 

To assess the significance of these movements, we compared the $\chi^2$ values obtained from zero-order ($y = \text{Const.}$) and first-order ($y = a\log E + b$) polynomial fits, where $y$ represents either $\alpha$ or $\delta$. The test statistic is:
\begin{equation}
    \text{TS} = \Delta \chi^{2} = \chi^{2}_{0} - \chi^{2}_{1},
\end{equation}
where \(\chi^2_0\) and \(\chi^2_1\) are the values from the 0th and 1st-order fits, respectively. The statistical significance is estimated by \(\sqrt{\text{TS}}\). For Region B, \(\chi^2_0 = 135.6\) and \(\chi^2_1 = 3.6\), resulting in a significance of 11.5\(\,\sigma\). For Region $\mathrm{\widetilde{D}}$, \(\chi^2_0 = 245.8\) and \(\chi^2_1 = 2.6\), yielding a significance of 15.6\(\,\sigma\). Thus, there is strong evidence of energy-dependent movement in Region $\mathrm{\widetilde{D}}$ and displacement in Region B. We also examined the energy-dependent movement of deficit regions and observed that Region III exhibits a significance of approximately \(3.7\,\sigma\), indicating a hint of energy-dependent shifting for one of the cold spots.

The average fractional relative intensity \cite{ARGO_MediumScale2013} is calculated as: 

\begin{equation}
\delta I_{\text{ROI}} = \frac{\sum_{i \in \text{ROI}} N_{i}}{\sum_{i\in \text{ROI}} B_{i}} -1,  
\end{equation}
where \(N_{i}\) is the observed event count in the \(i\)th pixel within the region of interest (ROI), and \(B_{i}\) is the estimated background count, including isotropic and low multipole contributions (\(\ell \leq 2\)). The size of ROIs for estimation of fractional relative intensity is chosen to be consistent with the most significant regions with over 7\,$\sigma$ (or below $-7\,\sigma$) in Figure \ref{fig:ts}, and for Regions B and $\mathrm{\widetilde{D}}$, results are derived from energy-dependent ROIs, with geometric centroids set as per Figure \ref{fig:Eposition}.

\begin{figure}[!htbp]
\centering
\includegraphics[width=0.45\textwidth]{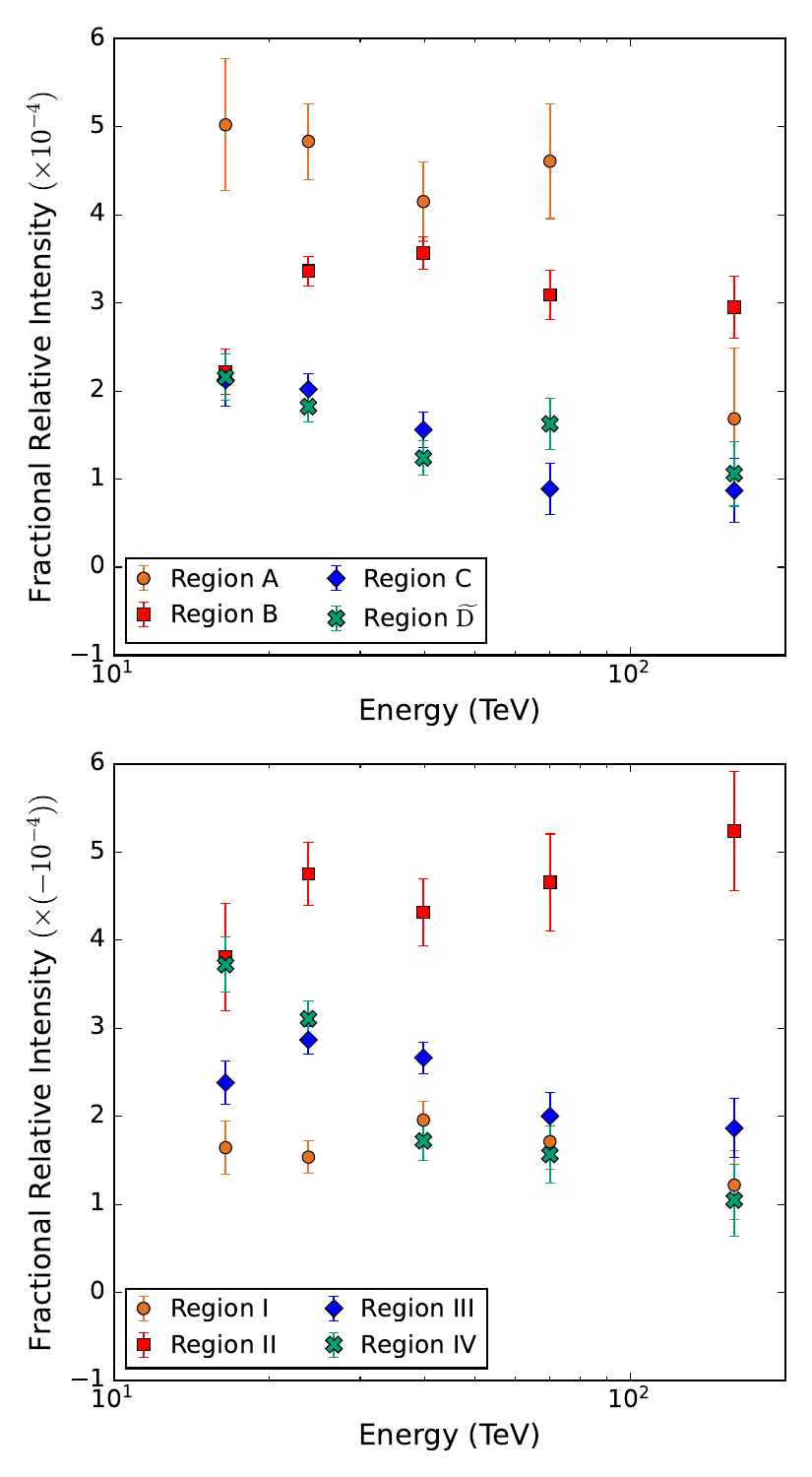}
\caption{Average fractional relative intensity of excess and deficit regions for each energy bin. The ROIs used to calculate the average fractional relative intensity are defined by a statistical significance of above $7\sigma$ for excess regions and below $-7\sigma$ for deficit regions. For Regions B and D, the definition of the ROIs is energy-dependent.}
\label{fig:deltaI}
\end{figure}

Figure \ref{fig:deltaI} illustrates the average fractional relative intensity of both excess and deficit regions. Regions B and $\mathrm{\widetilde{D}}$ exhibit flat spectra for excess structures, whereas Regions A and C show decreases with energy, with Region A decreasing more rapidly. Regarding the deficit regions, Regions I, II, and III exhibit relatively flat spectral features, whereas Region IV shows a monotonic decrease in intensity with increasing energy, potentially indicating a physical connection with the excess regions. The persistence of Regions A and B at around 100 TeV may pose challenges for heliospheric theories regarding their origin.
  
We compare the fractional relative intensities of Regions A and B with measurements from ARGO-YBJ \cite{ARGO_MediumScale2013} and HAWC \cite{HAWC_LSA2018}. While the amplitude of Region B exhibits excellent agreement across all observatories, the LHAASO measurements for Region A align more closely with ARGO-YBJ. The moderate level of disagreement with HAWC in Region A is attributed to differences in ROI selection and intensity extraction techniques. Furthermore, the energy-dependent morphological evolution within Region A corroborates the high-energy tail behavior previously reported by ARGO-YBJ. Detailed comparisons are provided in the Supplemental Material.

\noindent{\bf Discussion.}


Based on current measurements, we have sufficient statistical evidence to report that Regions B and $\mathrm{\widetilde{D}}$ exhibit energy-dependent shifts. Additional data are required to confirm the energy dependence of Region III. Furthermore, the remaining structures exhibit negligible energy-dependent evolution and remain effectively invariant across the observed energy range. According to \cite{Giacinti2012}, the energy-dependent evolution of medium-scale anisotropy position is attributed to the detailed turbulence of the Galactic magnetic field, particularly within the scattering length from Earth. Since CR scattering is influenced by the rigidity of CRs, the local volume of the turbulent field involved also becomes energy-dependent, which leads to changes in medium-scale anisotropies with energy.
The energy-dependent shifts observed in Regions B and $\mathrm{\widetilde{D}}$ provide important constraints on the structure of local magnetic fields. However, only specific field configurations can account for such shifts, thereby excluding others. In the supplementary materials, we present a simulated configuration where two medium-scale anisotropies exhibit similar shifts. A detailed investigation into the properties of the turbulent magnetic fields responsible for this energy-dependent displacement will be necessary in future work.

Regarding energy-independent structures, \cite{GiacintiKirk17} suggests that these may form in local interstellar turbulence if they are aligned with the local magnetic field direction and, consequently, with the large-scale anisotropy. Certain pitch-angle diffusion coefficients might create such structures, potentially constraining this critical yet poorly understood parameter. Region A, located near the direction of the dipole, might be a candidate for this type of formation. Previous full-sky anisotropy analyses, such as those conducted in \cite{IceCubeHAWC_2019, IceCubeHAWC_2025}, have established critical benchmarks for characterizing large-scale structures. Future analyses incorporating Region A, which bridges the Northern and Southern Hemispheres, will be instrumental in achieving a more integrated global mapping of cosmic-ray distributions.

Existing models primarily address Regions A and B. As these regions align with the direction of the helio-tail, it has been hypothesized that they may be influenced by the heliosphere. Related models consider particle acceleration through magnetic reconnection in the helio-tail and particle propagation in specific structures of the heliospheric electric and magnetic fields. However, these models do not account for Region $\mathrm{\widetilde{D}}$. 
It remains unclear whether other theoretical models—such as loss-cone leakage from a local source region via a magnetic trap \cite{MagneticMirror2008}, strong anisotropic Alfvén wave spectra induced by interstellar turbulence \cite{PreferrDirection2010}, or magnetic lens effects \cite{MagneticLens2015}—can account for the observed movement of Region $\mathrm{\widetilde{D}}$. Although these models incorporate energy-dependent scattering between cosmic rays and magnetic fields, further theoretical investigation is required to determine their consistency with our findings.

In addition to the energy-dependent movement of centroids, medium-scale anisotropy may show complex structures (e.g., Region C), and its morphological variations may encode rich information about the interactions between cosmic rays and magnetic fields. Moreover, angular power spectra also contain valuable information about turbulent magnetic fields and previous studies \citep{Ahlers_2014APS,Ahlers_2015APS,Zhang_2024,Bian_2025} have examined the roles of angular power spectra. Future research on morphology and angular power spectra from LHAASO-KM2A is expected to enhance our understanding of turbulent magnetic fields in the local environment.

\noindent{\bf Conclusion.}
In this study, we present measurements of medium-scale anisotropies in very high-energy CRs, ranging from 10 TeV to beyond 100 TeV, based on data collected by LHAASO-KM2A. Our analysis utilizes CR events accumulated from July 20, 2021, to July 19, 2024. We apply the all-distance equi-zenith angle method to estimate isotropic background reference maps. Subsequently, the relative intensity is expanded using spherical harmonic bases, and lower multipole moments with \(\ell \leq 2\) are subtracted to derive sky maps of medium-scale structures. In the resulting sky maps, we identify four statistically significant excess and four deficit regions.
Except for Region C, which is divided into two areas in the high-energy range, we investigated the central locations of the other excess and deficit regions for each energy bin, including both energy-independent and stable excess regions. Our findings provide strong evidence that the right ascension of Region $\mathrm{\widetilde{D}}$ is energy-dependent, while the declination of Region B also exhibits energy-dependent movement. This study is the first to report energy-dependent movements of medium-scale CR excess structures. Crucially, this dynamic evolution challenges existing static phenomenological models. It provides definitive evidence that these shifting structures are governed by the specific realization of the local turbulent Galactic magnetic field within the CR scattering length. Consequently, our findings transform medium-scale anisotropies from a static observational feature into a quantitative probe of the local interstellar magnetic environment.

Additionally, we report the detection of deficit regions and identify a possible hint of energy-dependent movement in one of these areas. The fractional relative intensity of medium-scale structures is measured beyond 100 TeV, revealing hard spectra for certain regions, where such spectral hardness is observed in both excess and deficit structures. LHAASO-KM2A's measurements provide new and valuable insights into CR propagation and the structure of the local turbulent environment. Further understanding of the turbulent Galactic magnetic field is anticipated with additional data from  LHAASO-KM2A and analyses extending to the sub-PeV energy range.

\noindent{\bf Acknowledgments.}
We would like to thank all staff members who work at the LHAASO site (CSTR: \href{https://cstr.cn/31117.02.LHAASO}{31117.02.LHAASO}) above 4400 meters above 
sea level year-round to maintain the detector and keep the water recycling system, electricity 
power supply and other components of the experiment operating smoothly. We are grateful to 
Chengdu Management Committee of Tianfu New Area for the constant financial support for research 
with LHAASO data. We deeply appreciate the computing and data service support provided by the 
National High Energy Physics Data Center for the data analysis in this paper. 
This research work is also supported by the following grants: the National SKA Program of China (2025SKA0110104), the National Natural Science Foundation of China NSFC (12273114, 12220101003), the Project for Young Scientists in Basic Research of the Chinese Academy of Sciences (YSBR-061), and Jiangsu Innovation and Entrepreneurship Talent Team Program (JSSCTD202436). In Thailand, support was provided by the National Science and Technology Development Agency (NSTDA), the National Research Council of Thailand (NRCT) under the High-Potential Research Team Grant Program (N42A650868).

\nocite{zhanghy2022}
\bibliography{citation}

\end{document}